\documentclass[twocolumn,showpacs,amsmath,amssymb]{revtex4}
\usepackage{graphicx}% Include figure files
\usepackage{dcolumn}% Align table columns on decimal point
\usepackage{bm}% bold math

\begin{document}

\preprint{APS/123-QED}

\title{Electronic phase separation in the pyrochlore double-exchange model}

\author{
Yukitoshi Motome$^1$ and
Nobuo Furukawa$^{2,3}$
}
\affiliation{%
$^1$Department of Applied Physics, University of Tokyo, Tokyo 113-8656, Japan
\\
$^2$Department of Physics and Mathematics, Aoyama Gakuin University, Kanagawa 229-8558, Japan
\\
$^3$Multiferroics Project, ERATO, Japan Science and Technology Agency (JST)
}

\date{\today}% It is always \today, today,

\begin{abstract}
Electronic phase separation and related inhomogeneity is 
ubiquitously seen in strongly-correlated systems. 
A typical example is found between ferromagnetic metal and 
antiferromagnetic insulator in CMR manganese oxides. 
Here we demonstrate that the geometrical frustration brings  
distinctive aspects into the phase separation phenomena. 
From Monte Carlo simulation and a simple energy comparison 
for the pyrochlore double-exchange model, 
we show that such phase separation takes place 
between ferromagnetic and paramagnetic metals. 
We discuss the relevance of our results 
to a spin-glassy metallic phase found in Mo pyrochlore oxides under external pressure. 
\end{abstract}

\maketitle

%%%%% Introduction

Electronic inhomogeneity is one of the key concepts in strongly correlated electron systems. 
It emerges from keen competition between different electronic phases, 
and gives rise to fascinating phenomena, 
such as nanometer-scale domain structure and gigantic response to external perturbations. 
Many intriguing topics related to the electronic inhomogeneity have been studied intensively, 
such as granular superconductivity in cuprates
\cite{Lang2002}, 
colossal magneto-resistance (CMR) in manganites
\cite{Dagotto2001}, and
relaxor response in dielectric compounds
\cite{Hirota2007}.

One of the most interesting, well-known examples is found in CMR manganites. 
It was argued that the competition 
between ferromagnetic metal (FM) and antiferromagnetic insulator (AFI) 
results in an electronically-inhomogeneous state 
in the presence of quenched disorder, and that 
the inhomogeneous state is highly sensitive to external magnetic field and gives rise to the CMR effect
\cite{Tokura2006}. 
Theoretically, the double-exchange (DE) model 
and its extensions have been extensively studied, 
and it was clarified that the model has an instability toward 
the electronic phase separation (PS)
\cite{Dagotto2001}. 
The PS occurs with a discontinuous change in the electron density, 
e.g., between FM and AFI: 
The former appears in a wide range of density 
by the DE ferromagnetic interaction of kinetic origin
\cite{Zener1951} and 
the latter is stabilized at a commensurate electron density by the AF super-exchange (SE) interaction
\cite{Yunoki1998,Moreo1999}. 
The energetically-competing phases lead to the electronic PS as well as bicritical behaviors, 
which are discussed as key players in the fascinating properties of CMR manganites. 

\begin{figure}[t]
 \centerline{\includegraphics[width=8.truecm]{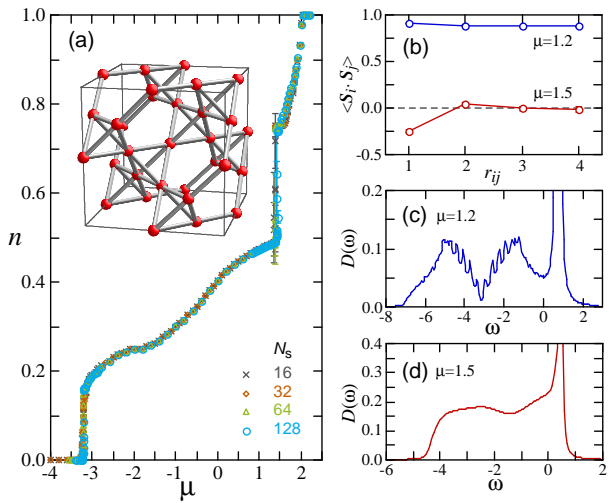}}
% \centerline{\includegraphics[width=12.truecm]{fig1.eps}}
\caption{
(Color online)
(a) Monte Carlo results for the electron density $n$ as a function of the chemical potential $\mu$ 
at $J_{\rm AF} = 0.04$ and $T=0.012$. 
The inset shows the cubic unit cell of the pyrochlore lattice. 
(b) Spin correlation $\langle {\mathbf S}_i \cdot {\mathbf S}_j \rangle$ 
for two states on the verge of the phase separation at $\mu_{\rm c1} \simeq 1.4$, 
plotted as a function of the distance in unit of the nearest-neighbor bond length. 
The data are measured along the 1D chains in the pyrochlore lattice. 
(c) and (d) Density of states for the two states. 
The data in (b)-(d) are for $N_{\rm s} = 128$ sites 
and the errorbar is in the symbol size or the line width. 
}
\label{fig:muvsn}
\end{figure}

In this Letter, we elucidate a new type of electronic PS in the DE model, 
which emerges when the lattice structure is geometrically frustrated. 
Here we show that PS takes place between FM and a {\it paramagnetic state} 
for the model defined on the pyrochlore lattice which has a three-dimensional network of 
corner-sharing tetrahedra [inset of Fig.~\ref{fig:muvsn}(a)]. 
The paramagnetic state does not have a charge gap, 
in contrast to the AFI state in the unfrustrated cases. 
While the PS was observed in the previous study 
\cite{Motome2010},
the present work elucidates its ubiquitousness and origin. 
It should be noted that this PS with a discontinuous change of the
density is non-trivial, since the transition from ferromagnet to
paramagnet is continuous in general.
This is in sharp contrast to the unfrustrated case where magnetic
transition between ferromagnet and antiferromagnet is necessarily of first order,
which can naturally accompany a jump in the electron density.

In addition to the new theoretical aspect, 
there is an experimental motivation for the frustrated DE models. 
It was proposed that the DE mechanism is relevant to FM state observed in 
Mo pyrochlore oxides $R_2$Mo$_2$O$_7$ ($R$ is rare earth)
\cite{Solovyev2003}. 
Recently, it was found that the compounds exhibit puzzling behaviors under external pressure
\cite{Iguchi2009}: 
FM is suppressed and taken over by a peculiar diffusive paramagnetic metal (PM). 
In the intermediate region, 
the system shows a spin-glassy metallic (SGM) behavior, 
which is in contrast with a spin-glassy insulating behavior observed in CMR manganites. 
The origin of SGM is not clear yet. 
We will discuss the relevance of our results to this SGM state.

%%%%% Model

Our model is the DE model 
defined on the pyrochlore lattice, shown in the inset of Fig.~\ref{fig:muvsn}(a). 
We consider the transfer integral and the SE coupling 
for nearest-neighbor (n.n.) sites only (the next n.n. hopping will be included later).
For simplicity, we take the limit of large Hund's-rule coupling and 
treat the localized spins ${\mathbf S}_i$ as classical vectors 
with a normalized length $|{\mathbf S}_i| = 1$. 
Then the model is effectively described by a spinless Hamiltonian in the form 
\cite{Anderson1955,MullerHartmann1996} 
\begin{equation}
{\cal H} = \sum_{\langle i,j \rangle} 
\big[ 
-\tilde{t}_{ij} (\tilde{c}_i^\dagger \tilde{c}_j + {\text{h.c.}})
+
J_{\text{AF}} \, {\mathbf S}_i \cdot {\mathbf S}_j
\big],
\label{eq:Htilde}
\end{equation}
where the first term describes the hopping of electrons 
whose spins are locally forced to align with the localized spins, 
and the second term represents the AF SE interaction between the localized spins. 
Here the effective transfer integral is given by
$
\tilde{t}_{ij} = t \, [ \cos (\theta_i/2) \cos (\theta_j/2) +
\sin (\theta_i/2) \sin (\theta_j/2) \exp \{ -i (\phi_i - \phi_j) \} ]
$,
where $(\theta_i, \phi_i)$ describes the angle of the localized spin ${\mathbf S}_i$. 
We set an energy unit as $t=1$ and the Boltzmann constant $k_{\rm B}=1$.

%%%%% Method

We calculate thermodynamic properties of the model (\ref{eq:Htilde}) 
by Monte Carlo (MC) simulation
\cite{Yunoki1998}, %. 
for the system sizes $N_{\rm s}=16$, $32$, $64$, and $128$ sites 
which correspond to $1 \! \times 1 \! \times \! 1$, 
$1 \! \times 1 \! \times \! 2$, $2 \! \times 2 \! \times \! 1$, and 
$2 \! \times 2 \! \times \! 2$ cubic unit cells of the pyrochlore lattice 
shown in the inset of Fig.~\ref{fig:muvsn}(a). 
To suppress the finite-size effect, we take an average over the twisted boundary conditions
\cite{Poilblanc1991,Gros1992,Motome_preprint} 
which is a well-established technique to smear out 
the extrinsic ``staircase-like" structure of the electron density 
as a function of the chemical potential.
This averaging is efficient, in particular, to identify PS region, since it enables us to extract 
the intrinsic jump associated with PS even in small-size clusters, as demonstrated below.

%%%%% Results

Figure~\ref{fig:muvsn}(a) shows a typical MC result for the electron density per site 
$n = \sum_i \langle \tilde{c}_i^\dagger \tilde{c}_i \rangle / N_{\rm s}$ 
as a function of the chemical potential $\mu$.
It is clearly seen that the system-size dependence 
is suppressed even at the low temperature ($T$), 
owing to the averaging over the twisted boundary conditions. 
The results reveal two sharp jumps: 
One is from $n \simeq 0.49$ to $n \simeq 0.75$ at $\mu_{\rm c1} \simeq 1.4$, and
the other is from $n \simeq 0.04$ to $n \simeq 0.16$ at $\mu_{\rm c2} \simeq -3.2$. 
These two jumps signal the electronic PS; 
the system is not stable at a density in these regions 
and phase-separated into two states at the edges of jump. 

Both the PS take place between ferromagnetic and paramagnetic states. 
This is corroborated by the spin correlation plotted in Fig.~\ref{fig:muvsn}(b). 
The results show typical behaviors on two sides of the PS at $\mu_{\rm c1} \simeq 1.4$. 
At $\mu < \mu_{\rm c1}$, 
the spin correlation is large and converges to a positive value for farther neighbors, 
indicating a long-range ferromagnetic ordering. 
On the other hand, at $\mu > \mu_{\rm c1}$, 
the spin correlation is AF for neighboring sites, but decays quickly to zero, 
suggesting a disordered paramagnetism. 

The ferromagnetic and paramagnetic states are both metallic. 
Figures~\ref{fig:muvsn}(c) and \ref{fig:muvsn}(d) present 
the density of states (DOS) per site for the two states. 
The results show that both states have no energy gap at the Fermi energy $\omega = 0$. 
This is consistent with the $\mu$ dependence of $n$ in Fig.~\ref{fig:muvsn}(a): 
$\partial n / \partial \mu$ is nonzero on the both sides of PS, namely, 
the system is charge compressible. 
The ferromagnetic state is stabilized by the DE interaction, and hence, it is metallic. 
The paramagnetic state is also considered to remain metallic; 
a gap formation by a simple AF ordering is suppressed by strong frustration 
\cite{note}. 
All the above situations of the magnetic and electronic states 
hold in a similar manner 
for PS at a lower density at $\mu_{\rm c2} \simeq -3.2$, 
while in this case the system is the ferromagnetic (paramagnetic) state 
for $\mu > \mu_{\rm c2}$ ($\mu < \mu_{\rm c2}$). 

\begin{figure}[t]
 \centerline{\includegraphics[width=8.5truecm]{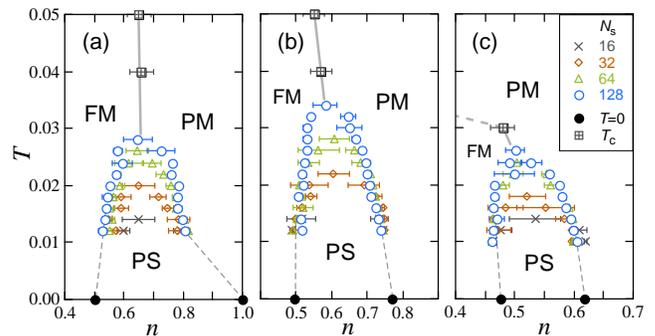}}
% \centerline{\includegraphics[width=12.truecm]{fig2.eps}}
\caption{
(Color online)
Phase diagrams determined by MC calculations 
at (a) $J_{\rm AF} = 0.00$, (b) $0.04$, and (c) $0.08$. 
Symbols surrounding the PS regions 
show the critical densities where $n$ changes discontinuously as a function of $\mu$. 
Filled circles at $T=0$ correspond to the phase boundaries in Fig.~\ref{fig:Ecompare}(b). 
Crossed squares indicate the second-order phase boundary between FM and PM. 
The lines are guides for the eye.
}
\label{fig:T_PS}
\end{figure}

We therefore conclude that the electronic PS occurs between the FM and PM states. 
This is a new type of PS appearing between metallic states, 
in sharp contrast with that in the unfrustrated case 
between FM and AFI with a clear energy gap~\cite{Yunoki1998,Moreo1999}. 
In the present model, we have shown in our previous paper~\cite{Motome2010} 
a crossover from the simple PM to the cooperative PM (CPM) states; 
the crossover temperature generally goes to zero on the verge of PS 
because of the cancellation between DE ferro and SE AF interactions. 
Therefore, the electronic PS discussed here should in general occur
above the crossover temperature and thus between FM and the simple PM.

The finite-$T$ phase diagram for PS is determined 
by keeping track of the discontinuity of $n(\mu)$. 
The results are shown for the higher-density PS at $\mu_{\rm c1}$ 
for different values of $J_{\rm AF}$ in Fig.~\ref{fig:T_PS}. 
The PS region appears in a dome-like shape in the plane of $n$ and $T$, 
whose top connects to the second-order phase boundary between FM and PM. 
As increasing $J_{\rm AF}$, the maximum temperature of PS is first enhanced slightly, 
but is suppressed for larger $J_{\rm AF}$: 
PS disappears for $J_{\rm AF} > 0.1$ in the $T$ range that we have calculated 
in MC ($T>0.01$). 

\begin{figure}[t]
 \centerline{\includegraphics[width=8.5truecm]{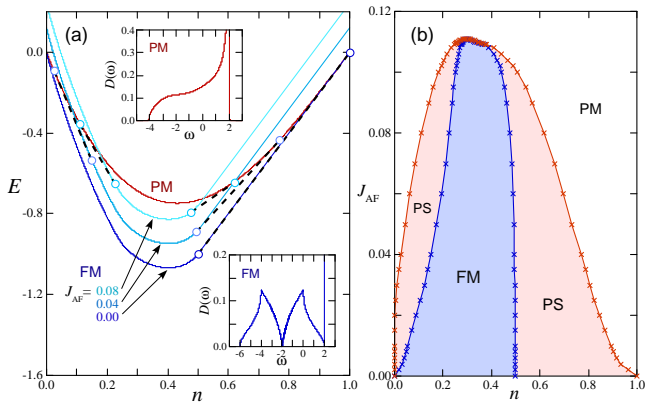}}
% \centerline{\includegraphics[width=12.truecm]{fig3.eps}}
\caption{
(Color online)
(a) Comparison of the ground-state energies between 
the completely-disordered PM and 
the perfectly-ordered FM. 
The dashed lines connecting the circles are common tangent lines for two energy curves, 
identifying the PS regions. 
The insets are the density of states for PM 
(upper inset) and FM (lower inset). 
(b) Phase diagram determined by the energy comparison. 
See the text for details.
}
\label{fig:Ecompare}
\end{figure}

Now we try to understand the origin of PS 
on the basis of a simple argument on the energy comparison. 
Here we assume that FM is in the perfectly-ordered (fully-saturated) state and 
PM is in the completely-disordered state 
which is approximately obtained by using PM solution of the dynamical
mean-field theory~\cite{Furukawa1994}.
In this assumption, we neglect not only the short-range AF correlations in the PM states, 
but also possibility of any other symmetry-broken states. 
We will see, however, that the energy comparison 
between the two simple states gives a reasonable estimate of PS regions 
in comparison with the finite-$T$ MC results.

Figure~\ref{fig:Ecompare}(a) shows the results for this analysis.
The lowest curve in the figure is the energy for the FM state at $J_{\text{AF}} = 0$, 
which is essentially given by that for non-interacting spinless fermions.
The FM state has two dispersive bands for $-6 \le \omega \le -2$ and $-2 \le \omega \le 2$, 
and two degenerate flat bands at $\omega = 2$. 
Because of the $\delta$-functional flat bands, 
the energy as a function of $n$ is given by a straight line 
from $n=0.5$ to $n=1.0$ in Fig.~\ref{fig:Ecompare}(a). 
DOS is shown in the inset with the one for the PM state. 
When we turn on $J_{\rm AF}$, the perfectly-ordered FM state costs the SE AF energy, 
though the completely-disordered PM state does not 
because the SE energy is averaged to be zero. 
Figure~\ref{fig:Ecompare}(a) illustrates the situation for $J_{\rm AF} = 0.04$ and $0.08$. 
We can identify the PS region by drawing common tangent lines 
for the energy curves of FM and PM states (dashed lines in the figure). 
Note that PS appears in both small and large density regions for $J_{\rm AF} > 0$. 

We carry out the energy comparison with varying $J_{\rm AF}$, 
and summarize the phase diagram shown in Fig.~\ref{fig:Ecompare}(b). 
In the middle of the phase diagram, we have the DE FM state, 
which shrinks as increasing $J_{\rm AF}$ because of the loss of kinetic energy. 
The FM state is surrounded by the PS region between FM and PM
\cite{note3}. 
The PS has a finite width in all regions and 
a direct continuous transition from FM to PM is prohibited, 
except for a marginal point at $(n, J_{\rm AF}) \simeq (0.3, 0.11)$. 

Let us compare the energy argument and the finite-$T$ MC results. 
The phase boundaries obtained from the analyses in Fig.~\ref{fig:Ecompare}
are plotted as the filled circles at $T=0$ in Fig.~\ref{fig:T_PS}. 
The finite-$T$ MC phase boundaries appear to be extrapolated smoothly to the $T=0$ results, 
except at $J_{\rm AF} = 0$, where some discrepancy in the right boundary is seen. 
It is probably due to the singular $\delta$-functional DOS 
discussed in Fig.~\ref{fig:Ecompare}(a). 
The overall agreement supports that our simple energy comparison is not far from the reality and 
that the PS is essentially a metal-to-metal one from FM to PM.

\begin{figure}[t]
 \centerline{\includegraphics[width=7.8truecm]{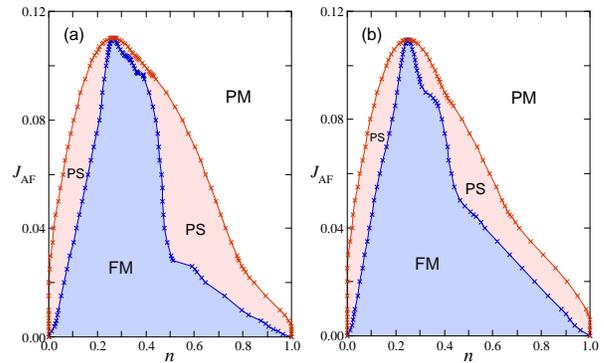}}
% \centerline{\includegraphics[width=12.truecm]{fig4.eps}}
\caption{
(Color online)
Phase diagram determined by the energy comparison 
for the model with next nearest-neighbor hopping $t'$ 
at (a) $t'=0.1$ and (b) 0.2.
}
\label{fig:t'}
\end{figure}

This peculiar PS is not specific to the present model (\ref{eq:Htilde}). 
We demonstrate it here by extending the model with including the next n.n. hopping $t'$. 
Figure~\ref{fig:t'} presents the results of the similar energy comparison for the extended model. 
For finite $t'$, the flat-band singularity in FM is removed, 
at the same time, the sharp cutoff at the band top in PM is rounded off 
[cf. the insets of Fig.~\ref{fig:Ecompare}(a)]. 
Nonetheless the metal-to-metal PS remains robust in a similar manner, 
as shown in the phase diagrams in Fig.~\ref{fig:t'}. 
PS will be robustly observed even when the model is extended 
to describe more complicated band structure and orbital degeneracy. 

We compare our results with those obtained in the absence of frustration.
In the absence of frustration, PS takes place between FM and AFI as mentioned in the introduction. 
AFI appears at a commensurate filling, such as $n=1.0$. 
The bandwidth of the AFI state is much smaller than that of FM, 
which drives the discontinuous change of the electron density associated with PS. 
Furthermore, the discontinuity is naturally expected 
since the magnetic transition between ferromagnet and antiferromagnet is necessarily of first order. 
In contrast to the unfrustrated case, 
our PS obtained for the frustrated pyrochlore model occurs between FM and PM. 
The severe frustration suppresses the AF long-range order.
In the ground state, the DE-induced ferromagnetism bears full saturation of moment 
to maximize the kinetic energy, and hence, 
if there is a transition from FM to a paramagnet by changing the chemical potential, 
the transition becomes discontinuous with a jump of the moment. 
The magnetic discontinuity naturally accompanies a discontinuous change of the electron density, 
which leads to PS, as demonstrated in Figs.~\ref{fig:Ecompare} and \ref{fig:t'}. 
The AF SE interaction enlarges the paramagnetism to doped metallic regions with reducing the FM region, 
resulting in the peculiar metal-to-metal PS in our model. 
Hence the complete polarization of the DE FM state at $T=0$ 
plays a key role in the present PS. 

We note that the argument is in some sense precipitate and 
the situation will be more complicated in the low-$T$ limit for large $J_{\rm AF}$ region, 
since the spin disordered state might be unstable toward some symmetry breaking,
such as an incommensurate magnetic ordering. 
It is anticipated in the present DE model that 
the kinetic motion of electrons, which in general leads to farther-neighbor DE interactions, 
acts as a degeneracy-lifting perturbation. 
In fact, in the pyrochlore Heisenberg AF spin system, 
it is known that any perturbation such as farther-neighbor exchange interactions 
may force the system to order 
\cite{Reimers1991}. 
What we have revealed is that 
the $T$ scale for such anticipated phase transitions is very small in the present DE model 
(at least, smaller than the $T$ range reached in the MC simulation) and 
the PS between FM and PM takes place well above it. 
We expect that a similar PS is widely seen in 
the DE systems on other frustrated lattice structures. %, 

Let us discuss the relevance of our results to experiments. 
Mo pyrochlore oxides $R_2$Mo$_2$O$_7$ exhibit peculiar phase transitions under external pressure
\cite{Iguchi2009}. 
FM becomes unstable and is taken over by a diffusive PM; in between, 
a spin-glassy metallic (SGM) state appears at low $T$. 
In the transition to SGM, there is a glassy response in the AC susceptibility 
but no substantial anomaly in the resistivity. 
This SGM is peculiar because a spin-glassy behavior is observed 
in the insulating phase in the case of CMR manganites. 
The spin-glassy insulating behavior can be understood 
by competition between FM and AFI with 
a formation of FM clusters in the matrix of AFI (or charge/orbital ordered insulator). 
In the Mo pyrochlores, the competing phases under pressure are FM and PM, 
and therefore, we speculate that the intervening SGM state originate from
the electronic PS found in our results. 
The glassy magnetic response is presumably explained 
by domain-like structure or a mixed state; 
for such situation, it is crucial to consider the effect of long-ranged Coulomb interaction 
as well as quenched randomness~\cite{Sboychakov2007,Shenoy2009}. 
In the present case, the competing phases are both metallic, and hence, 
we expect that the resulting glassy state remains metallic. 
Further study by extending our model is necessary to confirm this scenario. 
It is also interesting to extend the experimental study away from the commensurate filling 
for comprehensive understanding of the phenomena. 

Finally we comment on the relation between PS and the flat band singularity at $J_{\rm AF} = 0$. 
It was argued that the DE FM is no longer stable 
when the chemical potential is in the flat band 
\cite{Penc1999}. 
This instability corresponds to the jump from FM at $n=0.5$ to 
the macroscopically-degenerate state at $n=1.0$ 
in Fig.~\ref{fig:Ecompare}. 
A jump of electron density associated with the flat band singularity 
smoothly evolves to PS for $J_{\rm AF} > 0$. 
This suggests a possibility to understand the instability 
from the viewpoint of the electronic PS. 
Further analysis is left for future study.

%%%%% Summary

To summarize, we have investigated the electronic phase separation 
in the double-exchange model defined on the frustrated pyrochlore lattice. 
The phase separation is found to occur between the ferromagnetic metal and 
the paramagnetic metal: The former is stabilized by the double-exchange interaction, and 
the latter is induced by the frustrated antiferromagnetic super-exchange interaction. %, 
We discuss that the phase separation explains the peculiar spin-glassy metallic behavior 
in Mo pyrochlore oxides under external pressure. 
The metal-to-metal phase separation is characteristic of the frustrated double-exchange system, 
which is not seen in the unfrustrated models studied for the colossal magneto-resistive manganites.

%%%%% Acknowledgment

We would like to thank S. Iguchi and Y. Tokura for stimulating discussions 
on their experimental results.
Y. M. appreciate fruitful discussions with K. Penc and Y. Yamaji. 
This work was supported by Grants-in-Aid for Scientific research 
(Nos. 17071003, 19052008, and 21340090), 
by Global COE Program ``the Physical Sciences Frontier", and 
by the Next Generation Super Computing Project, Nanoscience Program, MEXT, Japan.


\begin{thebibliography}{9}
\bibitem{note}
Another possibility is the Anderson insulator, 
but it is unlikely in the clean system without randomness.
\end{thebibliography}
\end{document}